# Band-selective simulation of photoelectron intensity and converging Berry phase in trilayer graphene

Hayoon Im[a], Sue Hyeon Hwang[a], Minhee Kang[a], Kyoo Kim[b], Haeyong Kang[a,c],*, and Choongyu Hwang[a,c],*

[a]Department of Physics, Pusan National University, Busan 46241, Republic of Korea

[b]Korea Atomic Energy Research Institute, Daejeon 34057, Republic of Korea

[c]Quantum Matter Core-Facility, Pusan National University, Busan 46241, Republic of Korea

**Abstract**

Berry phase is one of the key elements to understand quantum-mechanical phenomena such as the Aharonov-Bohm effect and the unconventional Hall effect in graphene. The Berry phase in monolayer and bilayer graphene has been manifested by the anisotropic distribution of photoelectron intensity along a closed loop in the momentum space as well as its rotation by a characteristic angle upon rotating light polarization. Here we report the band-selective simulation of photoelectron intensity of trilayer graphene to understand its Berry phase within the tight-binding formalism. ABC- and ABA-stacked trilayer graphene show characteristic rotational angles of photoelectron intensity distribution, as predicted from their well-known Berry phases. Surprisingly, however, in ABA-stacked trilayer graphene, the rotational angle changes upon approaching toward the band touching point between the conduction and valence bands, which suggest that Berry phase changes as a function of binding energy. The binding energy-dependent Berry phase is attributed to the enhanced hybridization of the two electron bands of ABA-stacked trilayer graphene that converge at the band touching point, resulting in the converging Berry phase. These findings will provide an efficient way of tuning Berry phase and hence exotic phenomena stemming from the Berry phase.

*Corresponding author.

Email: haeyong.kang@pusan.ac.kr, ckhwang@pusan.ac.kr

# 1. Introduction

Berry phase is a geometric phase that a quantum state acquires when it undergoes adiabatic cyclic evolution along a closed loop in a parameter space [1,2], which is responsible for novel physical phenomena such as the Aharonov-Bohm effect [3-5] and the unconventional Hall effect [6,7]. Berry phase has been extensively studied to understand topological properties of quantum materials including topological insulators [8,9], spin-orbit coupled systems [10-12], and transition-metal dichalcogenides [13,14]. Graphene is also a prototypical quantum material with well-known Berry phase [6,7,15-19]. Due to the characteristic honeycomb lattice of graphene, quasiparticles in graphene close to the band touching point between the conduction and valence bands around the Brillouin zone corner, K point, are approximately described by spinor eigenstates that give the pseudospin nature to the quasiparticles. The pseudospin is locked to the momentum in graphene, directly revealing the Berry phase of quasiparticles in graphene. In other words, when the pseudospin travels along a closed loop in the momentum space, $n$-layer graphene ($n=1$ for monolayer, $n=2$ for bilayer) exhibits a pseudospin winding number of $n$, corresponding to $n\pi$ Berry phase [20-22]. The Berry phase in graphene leads to the anisotropic distribution of angle-resolved photoemission spectroscopy (ARPES) intensity, which rotates by a characteristic angle when light polarization used to excite quasiparticle is rotated by $\pi/2$ [16,18,19,23]. The $\pi/n$ rotation of ARPES intensity distribution for $n$-layer graphene originates from the $n\pi$ Berry phase [18].

Despite extensive experimental and theoretical studies on graphene, previous studies on Berry phase have mainly focused on monolayer and bilayer graphene, while trilayer graphene have been rarely studied so far [24-30]. Depending on the stacking order, trilayer graphene has rhombohedral or Bernal structures, so-called ABC- or ABA-stacked graphene, whose crystalline structures are shown in Figs. 1(a)-(d). A and B sublattices of each graphene layer are colored by dark and bright balls, when bottom, middle, and top graphene layers are denoted by 1, 2, and 3, respectively. Each trilayer graphene is theoretically predicted to exhibit a Berry phase of $3\pi$ and $2\pi + \pi$ [21], different from the uniquely defined Berry phases of monolayer and bilayer graphene.

Here we report the simulation of ARPES intensity for trilayer graphene with different stacking orders within the tight-binding formalism and explore the characteristics of the Berry phase via the band-selective analysis. When light polarization is rotated by π/2, the ARPES intensity distribution of all the bands close to the band touching point of ABC-stacked trilayer graphene rotates by π/3. On the other hand, two different rotational angles of about π/2 and π are observed for bands of ABA-stacked trilayer graphene. These results indicate the Berry phases of 3π and 2π + π for ABC- and ABA-stacked graphene, respectively, consistent with the prediction based on the previous work on monolayer and bilayer graphene [18] and the Berry phase of trilayer graphene [21]. Surprisingly, however, in ABA-stacked trilayer graphene, the two different rotational angles of the monolayer- and bilayer-like bands converge to about π/2 upon approaching toward the band touching point, suggesting about 2π Berry phase, instead of the 2π + π Berry phase. The change in the rotational angle as a function of binding energy is attributed to the hybridized wave function of the two different bands and the resultant convergence of Berry phase. These findings will provide an important insight on manipulating not only Berry phase in condensed matters, but also exotic phenomena arising from the Berry phase.

## 2. Methods

To simulate ARPES intensity, the Hamiltonian was considered using the tight-binding model for the $p_z$ orbital of each carbon with hopping parameters, $\gamma_0$, $\gamma_1$, $\gamma_3$, and $\gamma_4$, as depicted in Figs. 1(c)-(d). Each hopping parameter corresponds to intralayer nearest-neighbor ($A_i \leftrightarrow B_i$ for i={1,2,3}) hopping, interlayer nearest-neighbor interlayer ($B_i \leftrightarrow A_{i+1}$ for i={1,2}), and interlayer hopping between different sublattices ($A_i \leftrightarrow B_{i+1}$ for i={1,2}) and between the same sublattices ($A_i \leftrightarrow A_{i+1}$ and $B_i \leftrightarrow B_{i+1}$ for i={1,2}), respectively. The values of each parameter used in the simulations are $\gamma_0 = -3.1$ eV, $\gamma_1 = 0.38$ eV, and $\gamma_4 = -0.141$ eV [35]. $\gamma_3$ was chosen to be zero to minimize the trigonal wrapping effect. The basis set composed of Block sums of localized orbitals on each sublattice was introduced: $\phi_1 = (A_1 - A_3)/\sqrt{2}$, $\phi_2 = (B_1 - B_3)/\sqrt{2}$, $\phi_3 = (A_1 + A_3)/\sqrt{2}$, $\phi_4 = B_2$, $\phi_5 = A_2$, and $\phi_6 = (B_1 + B_3)/\sqrt{2}$. The potential difference between two outer layers was introduced to be $\Delta_1 = 0.05$ eV, and that of the two outer layers with respect to the middle layer was $\Delta_2 = -0.023$ eV, while the potential difference between $B_1$ and $A_3$ sites was set to be $\delta = -0.0105$ eV [35]. Within this setup, the tight-binding Hamiltonians for ABC- and ABA-stacked trilayer graphene are as follows [21,32-36].

$$H_{ABC} = \begin{pmatrix} \Delta_1 + \Delta_2 & \gamma_0 u(\vec{k}) & \gamma_4 u(\vec{k}) & \gamma_3 u^*(\vec{k}) & 0 & 0 \\ \gamma_0 u^*(\vec{k}) & \Delta_1 + \Delta_2 & \gamma_1 & \gamma_4 u(\vec{k}) & 0 & 0 \\ \gamma_4 u^*(\vec{k}) & \gamma_1 & -2\Delta_2 & \gamma_0 u(\vec{k}) & \gamma_4 u(\vec{k}) & \gamma_3 u^*(\vec{k}) \\ \gamma_3 u(\vec{k}) & \gamma_4 u^*(\vec{k}) & \gamma_0 u^*(\vec{k}) & -2\Delta_2 & \gamma_1 & \gamma_4 u(\vec{k}) \\ 0 & 0 & \gamma_4 u^*(\vec{k}) & \gamma_1 & \Delta_2 - \Delta_1 + \delta & \gamma_0 u(\vec{k}) \\ 0 & 0 & \gamma_3 u(\vec{k}) & \gamma_4 u^*(\vec{k}) & \gamma_0 u^*(\vec{k}) & \Delta_2 - \Delta_1 + \delta \end{pmatrix}$$

$$H_{ABA} = \begin{pmatrix} \Delta_1 + \Delta_2 & \gamma_0 u(\vec{k}) & \gamma_4 u(\vec{k}) & \gamma_3 u^*(\vec{k}) & 0 & 0 \\ \gamma_0 u^*(\vec{k}) & \Delta_1 + \Delta_2 & \gamma_1 & \gamma_4 u(\vec{k}) & 0 & 0 \\ \gamma_4 u^*(\vec{k}) & \gamma_1 & -2\Delta_2 & \gamma_0 u(\vec{k}) & \gamma_4 u^*(\vec{k}) & \gamma_1 \\ \gamma_3 u(\vec{k}) & \gamma_4 u^*(\vec{k}) & \gamma_0 u^*(\vec{k}) & -2\Delta_2 & \gamma_3 u(\vec{k}) & \gamma_4 u^*(\vec{k}) \\ 0 & 0 & \gamma_4 u(\vec{k}) & \gamma_3 u^*(\vec{k}) & \Delta_2 - \Delta_1 + \delta & \gamma_0 u(\vec{k}) \\ 0 & 0 & \gamma_1 & \gamma_4 u(\vec{k}) & \gamma_0 u^*(\vec{k}) & \Delta_2 - \Delta_1 + \delta \end{pmatrix}$$

Here, $u(\mathbf{k}) = \sum_{i=1}^{3} \exp(i\mathbf{k} \cdot \mathbf{b}_i) = e^{\frac{ik_y 4\pi}{3\sqrt{3}}}(1 + 2\cos(\frac{k_x 2\pi}{3})e^{-\frac{ik_y 2\pi}{\sqrt{3}}})$, constructed with the three vectors in the real space connecting nearest neighbor carbon atoms $\mathbf{b}_1 = b(0,1)$, $\mathbf{b}_2 = b\left(-\frac{\sqrt{3}}{2}, -\frac{1}{2}\right)$, and $\mathbf{b}_3 = b\left(\frac{\sqrt{3}}{2}, -\frac{1}{2}\right)$, when $\mathbf{k}$ the vector representation in the momentum space and $b = 1.42$ Å the in-plane inter-carbon distance. The interaction Hamiltonian $H_{\text{int}}(\mathbf{k}, \mathbf{Q})$ coupling to the electromagnetic waves with a wave vector $\mathbf{Q}$ is defined as $-\frac{e}{c}\hat{\mathbf{A}} \cdot \hat{\mathbf{v}}$, where $\hat{\mathbf{A}}$ is the external vector potential and $\hat{\mathbf{v}} = [\hat{\mathbf{r}}, H]/i\hbar$ with $\hat{\mathbf{r}} = i\hbar(\nabla_{\mathbf{k}}, \partial k_z)$, $\hbar$ the Planck's constant, $e$ the charge of an electron, and $c$ is the speed of light [18]. Based on this setup, the photoelectron intensity has been calculated by the absolute square of the transition matrix element $M_{s\mathbf{k}} = \langle f_{\mathbf{k}+\mathbf{Q}}|H_{\text{int}}(\mathbf{k}, \mathbf{Q})|\psi_{s\mathbf{k}}\rangle$, where $|\psi_{s\mathbf{k}}\rangle$ is the eigenstate with $s = \pm 1$ the band index, $|f_{\mathbf{k}+\mathbf{Q}}\rangle$ the plane-wave final state projected onto the $p_z$ orbitals of graphene [18].

## 3. Results and discussion

Figures 1(e) and (f) show calculated electron band structures close to the band touching point around the K point for ABC- and ABA-stacked trilayer graphene using the full TB Hamiltonians discussed above. The ABC-stacked graphene consists of three parabolic bands, of which outermost conduction and valence bands touch slightly off the K point, while the two inner bands cross at about $\pm 0.4$ eV from the Fermi energy, $E_F$. The basis of ABA-stacked trilayer graphene that has mirror reflection symmetry along the out-of-plane direction can be recombined and block-diagonalized to [L1+L3, L2] (L=A, B) with mirror eigenvalue = +1 and [L1-L3] with eigenvalue = –1, resulting in the combination of bilayer-like parabolic bands ($M_z$ = +1) and monolayer-like linear bands ($M_z$ = –1). The outer-most parabolic band and the linear band merge together as they approach toward $E_F$.

To investigate the Berry phase of the trilayer graphene, ARPES intensity distribution was simulated within the TB formalism [18]. Figures 2(a) and (b) show the results simulated at 1.0 eV below $E_F$ for ABC- and ABA-stacked graphene, respectively. The white-dashed hexagon is the first Brillouin zone of graphene. For both constant energy contours, several crescent-like spectral features are observed around each K point. To examine this spectral feature in more detail, Figs. 2(c) and (d) show polarization-dependent ARPES intensity distribution simulated at –1.0, 0.0, and 1.0 eV with respect to $E_F$ around the K point indicated by the white squares in Figs. 2(a) and (b). The left and right panels are results obtained using light polarization parallel to the $k_x$ (X-polarization or X-pol.) and $k_y$ (Y-polarization or Y-pol.) axes of graphene, respectively. At $E_F$ where the conduction and valence bands touch each other, both X- and Y-polarization data exhibit a point-like shape, whereas upon moving away from $E_F$, the concentric crescent-like shapes appear, consistent with the characteristic conical electron band structure of trilayer graphene as shown in Figs. 1(e) and (f). In addition, the intensity distribution of the conduction band for X-polarization is almost the same as that of the valence band for Y-polarization. This originates from the chiral nature of quasiparticles in graphene [18]. Apart from these common features, overall intensity distribution of ABC-stacked trilayer graphene is different from that of ABA-stacked trilayer graphene.

To better understand the anisotropic intensity distribution of the three bands of each graphene, Figs. 2(e) and (f) show the band-selective intensity distribution for the inner, middle, and outer bands at $E - E_F = -1.0$ eV for X- and Y-polarization, respectively. Each band can be displayed separately by setting the broadening effect used in the simulation to zero. The inner band of the ABC-stacked trilayer graphene for X-polarization has maximum intensity along the K-Γ direction, whereas it lies along the K-M direction for Y-polarization. The intensity distribution of the middle and outer bands shows similar polarization dependence as the inner one, i.e., maximum intensity along the K-M (Γ-K) direction for X-polarization changes its direction to the K-Γ (K-M) direction for the middle (outer) band. On the other hand, ABA-stacked trilayer graphene exhibits slightly different polarization dependence from the ABC-stacked graphene. Despite the inner and outer bands show similar behavior as those of ABC-stacked graphene, the intensity maxima of the inner (outer) band for X-polarization (Y-polarization) is slight off the K-M (K-Γ) direction. More interestingly, for the middle band, the intensity maximum along the Γ-K direction changes its direction to the K-M direction upon changing light polarization, that is observed neither for the ABC-stacked trilayer graphene nor for the other two bands of ABA-stacked trilayer graphene.

The change in the ARPES intensity distribution can be quantitatively analyzed by taking the intensity profile for each band around the K point as shown in Fig. 3. The intensity profile is taken at $E - E_F = -1.0$ eV where $\theta$ is the angle with respect to the $+k_x$ axis as shown in the inset. The ARPES intensity of the inner, middle, and outer bands is plotted with black-solid, red-dashed, and yellow-solid curves, respectively. In ABC-stacked trilayer graphene, the intensity maxima of all the bands rotate by about $\pi/3$ upon rotating light polarization by $\pi/2$. On the other hand, in ABA-stacked trilayer graphene, the intensity maximum of the middle band rotates by about $\pi$, whereas the inner and outer bands rotate by about $\pi/2$.

For ABC- and ABA-stacked trilayer graphene, the spinor eigenstates are analytically obtained when the momentum with respect to the K point is assumed to be very small, allowing us to determine the pseudospin texture of each graphene, as shown in Fig. 3(c) [21,22]. The pseudospin of ABC-stacked graphene denoted by blue arrows in the upper panel shows a winding number of 3 when traveling along a closed loop in the momentum space. This indicates a Berry phase of $3\pi$, which is also applied to the other two bands of ABC-stacked graphene [21,24-26,33-37]. The pseudospin of ABA-stacked trilayer graphene exhibits different winding numbers of 1 and 2 for the linear (middle) and parabolic (outer and inner) bands as denoted by green and red arrows, which gives of $\pi$ and $2\pi$ Berry phase, respectively [21,25,26,33].

The rotation of ARPES intensity distribution in graphene upon rotating light polarization is explained by the Berry phase effect in monolayer and bilayer graphene [18]. The spectral intensity of $n$-layer graphene ($n$=1 for monolayer, $n$=2 for bilayer) rotates by $\pi/n$ due to $n\pi$ Berry phase when the light polarization is rotated by $\pi/2$ [18]. In the same analogy, the rotational angle of the ARPES intensity distribution is predicted to be $\pi/3$ for ABC-stacked trilayer graphene, whereas it is $\pi/2$ and $\pi$ for the parabolic and linear bands for ABA-stacked trilayer graphene, due to their Berry phases of $3\pi$ and $2\pi + \pi$, respectively. This prediction is manifested roughly in the simulations for the ARPES intensity distribution for trilayer graphene shown in Figs. 2 and 3.

The rotational angle of the outer and inner bands of ABA-stacked graphene, however, is slightly deviate from $\pi/2$. This discrepancy is likely due to the asymmetric line-shape of the intensity profile, which may distort angular information. Interestingly, the asymmetry of the line-shape is more pronounced in the middle band plotted by the red-dashed curve in the lower panel of Fig. 3(b). For Y-polarization, despite the spectral intensity has its maximum at 0 or $2\pi$, a clear shoulder of the intensity appears at about $\pi/2$ or $3\pi/2$. The unexpected spectral intensity suggests the possibility of additional effect that the isolated monolayer and bilayer graphene bands and their Berry phases cannot describe.

To find the origin of the asymmetric line-shape, ARPES intensity distribution was simulated for ABA-stacked trilayer graphene as a function of energy relative to $E_F$ as shown in Fig. 4. The position of the intensity maxima of the inner and outer bands do not notably change as denoted by white arrows in Figs. 4(a) and (b). On the other hand, the intensity maxima of the middle band for Y-polarization dramatically change their positions as denoted by white arrows in Fig. 4(c), while they remain the same for X-polarization. The rotational angle of the intensity maxima as a function of $E - E_F$ is summarized in Fig. 4(d) for the inner, outer, and middle bands with black, yellow and red symbols, respectively. The rotational angle of the inner band (black symbol) above $E - E_F = -0.7$ eV cannot be extracted as the valence band maximum of the inner band is about $-0.6$ eV as shown in Fig. 1(f).

At higher binding energies, e.g., $E - E_F = -1.1$ eV, the rotation angle of the inner, outer, and middle bands are $0.67\pi$, $0.61\pi$, and $\pi$, respectively. While they are similar to $0.5\pi$, $0.5\pi$, and $\pi$ that are predicted from the eigenstates obtained by the simple approximation of $|k - K| \ll 1$ [21], these results suggest that the Berry phase of each of the parabolic band of the ABA-stacked trilayer graphene is $1.5\pi$ and $1.65\pi$, and that of the linear band is $\pi$, following the analogy obtained from the monolayer and bilayer [18], and from ABC-stacked trilayer graphene in this study. Upon changing $E - E_F$, the rotational angle of the inner band remains almost the same suggesting that it has a robust Berry phase of $1.5\pi$. On the other hand, the rotational angle of the outer band gradually decreases from $0.61\pi$ at $E - E_F = -1.1$ eV to $0.52\pi$ at $E - E_F = -0.3$ eV. Surprisingly, that of the middle band abruptly changes above $E - E_F = -0.8$ eV until it goes down to $0.52\pi$ at $E - E_F = -0.3$ eV, which is the same rotational angle as that of the outer band. The equivalent rotational angle of the outer and middle bands suggests that the Berry phases of both bands converge to $1.9\pi$, that is very close to the Berry phase of $2\pi$ predicted from the bilayer-like outer band of the ABA-stacked trilayer graphene. The difference of the outer and middle bands compared to the inner band is that the latter remains intact independent on the energy relative to $E_F$, whereas the former two merge together upon decreasing $|E - E_F|$ as shown in Fig. 1(f). The merging of the two bands close to $E_F$ can lead to the enhanced hybridization between quasiparticles from the two bands, giving rise to the single quantum mechanical phase for the quasiparticles from two different bands. Meanwhile, due to non-zero $\Delta_1$, $\Delta_2$, and $\delta$, $H_{ABA}$ has non-zero off-diagonal terms when block diagonalized by basis transformation, leading the hybridization between the parabolic and linear bands. As a result, the Berry phases of the inner, middle, and outer bands deviate from those of unperturbed bands with zero $\Delta_1$, $\Delta_2$, and $\delta$, and converge around the band mering point between the parabolic and linear bands, as manifested in Fig. 4(d). One can also notice that the intensity profile of the ABC trilayer graphene shown in Fig. 3(a) also shows slight asymmetry, e.g., middle band for Y-polarization. Since the rotational angle of the intensity maxima of this system is well described by the Berry phase effect of

π/3, the asymmetry can also have additional contribution other than the merging band effect discussed above, such as the trigonal wrapping of the electron band structure, that can also be applied to the ABA trilayer graphene.

These results not only show that the band-selective Berry phase can be investigated via the simulation of ARPES intensity distribution within the TB formalism, but also suggest that Berry phase can be controlled through the hybridization of electron bands. Such an efficient way of tuning Berry phase can provide a plausible methodology to manipulate intriguing physical phenomena stemming from Berry phase such as the Aharonov-Bohm effect [3-5], the unconventional Hall effect [6,7], and the topological properties of topologically protected surface states [8,9].

## 4. Conclusions

The band-selective ARPES intensity distribution of trilayer graphene with different stacking structures has been simulated within the tight-binding formalism. For ABC-stacked trilayer graphene, when the light polarization is rotated by $\pi/2$, the photoelectron intensity rotates by $\pi/3$, which is consistent with the theoretical prediction of the $3\pi$ Berry phase. On the other hand, for ABA-stacked trilayer graphene, although the ARPES intensity distribution is expected to rotate by $\pi$ and $\pi/2$ due to the $\pi$ and $2\pi$ Berry phases, the outer and middle bands rotate by about $\pi/2$ close to $E_F$, suggesting a single Berry phase of about $2\pi$. The rotational angle gradually approaches toward the theoretically predicted values corresponding the $\pi$ and $2\pi$ Berry phases to as the electron band structures move away from $E_F$. This unusual spectral feature is attributed to the hybridization between quasiparticles from the outer and middle bands, that is enhanced by their mergence upon approaching toward $E_F$.

**Acknowledgements**

We gratefully acknowledge J. Hwang and H. C. Park for helpful discussion. This work was supported by the 2-Year Research Grant from the Pusan National University.

**Conflicts of Interest**

The authors declare no conflicts of interest.

**Figure Captions**

Figure 1. (a, b) Top views of (a) ABC- and (b) ABA-stacked trilayer graphene, where green, red, and blue balls represent carbon atoms in the bottom, middle, and top layers, denoted by 1, 2, and 3, respectively. A and B denote carbon sublattices in each graphene layer that are colored by dark and light balls. (c, d) Side views of (a) ABC- and (b) ABA-stacked trilayer graphene. Dashed arrows indicate tight-binding hopping parameters of trilayer graphene, denoted by $\gamma_0$, $\gamma_1$, $\gamma_3$, and $\gamma_4$. (e, f) Calculated electron band structure of (e) ABC- and (f) ABA-stacked trilayer graphene along the Γ-K-M direction using the tight-binding model.

Figure 2. (a, b) Simulated constant energy contours of (a) ABC- and (b) ABA-stacked trilayer graphene at $E - E_F = -1.0$ eV for the first Brillouin zone denoted by the white-dashed hexagon. The simulation has been done for X-polarized light. (c, d) Simulated constant energy contours of (c) ABC- and (d) ABA-stacked trilayer graphene for X-(left) and Y-(right) polarized light for the area denoted by the white rectangle in panels (a) and (b). (e, f) The inner (left panels), middle (middle panels), and outer (right panels) bands of (e) ABC- and (f) ABA-stacked trilayer graphene simulated with X- and Y-polarized lights at $E - E_F = -1.0$ eV.

Figure 3. (a, b) Angle-dependent intensity profiles of (c) ABC- and (d) ABA-stacked trilayer graphene extracted from Figs. 2(e) and 2(f). Black-solid, red-dashed, and yellow-solid curves are for inner, middle, and outer bands, respectively. The rotational angle $\theta$ is defined as the angle with respect to the $+k_x$ axis as shown in the inset. (c, d) Pseudospin textures of (c) ABC- and (d) ABA-stacked trilayer graphene.

Figure 4. (a-c) Constant energy contours of the (a) inner, (b) outer, and (c) middle bands of ABA-stacked trilayer graphene at several different energies with respect to $E_F$. White arrows denote the position of maximum spectral intensity. (d) Rotational angle of the maximum intensity upon rotating light

polarization by π/2 as a function of $E - E_F$ in the upper panel and band-selective Berry phase as a function of $E - E_F$ converted from the rotational angle in lower panel. Black, red, and yellow colors denote the inner, middle, and outer bands, respectively.

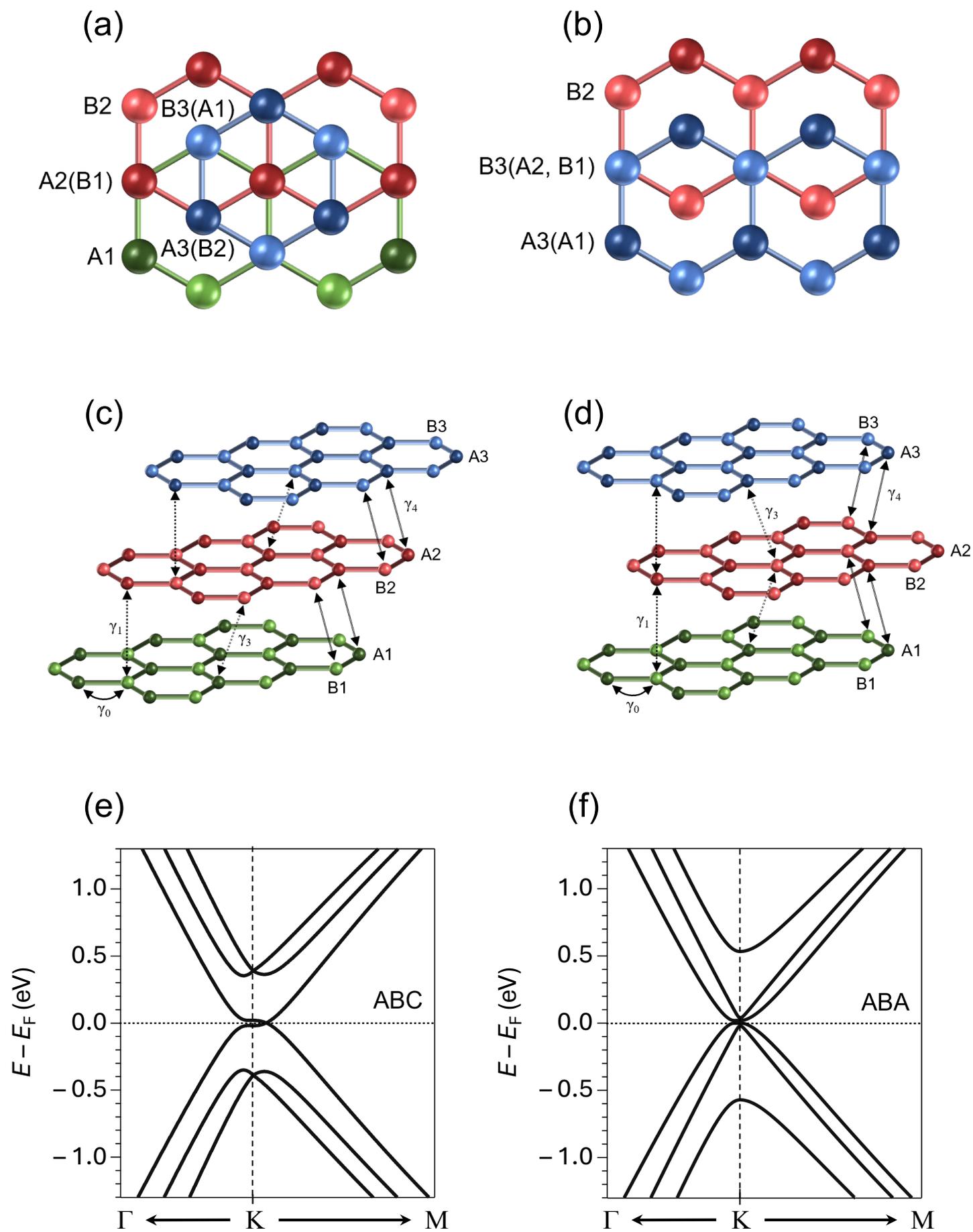

Fig. 1

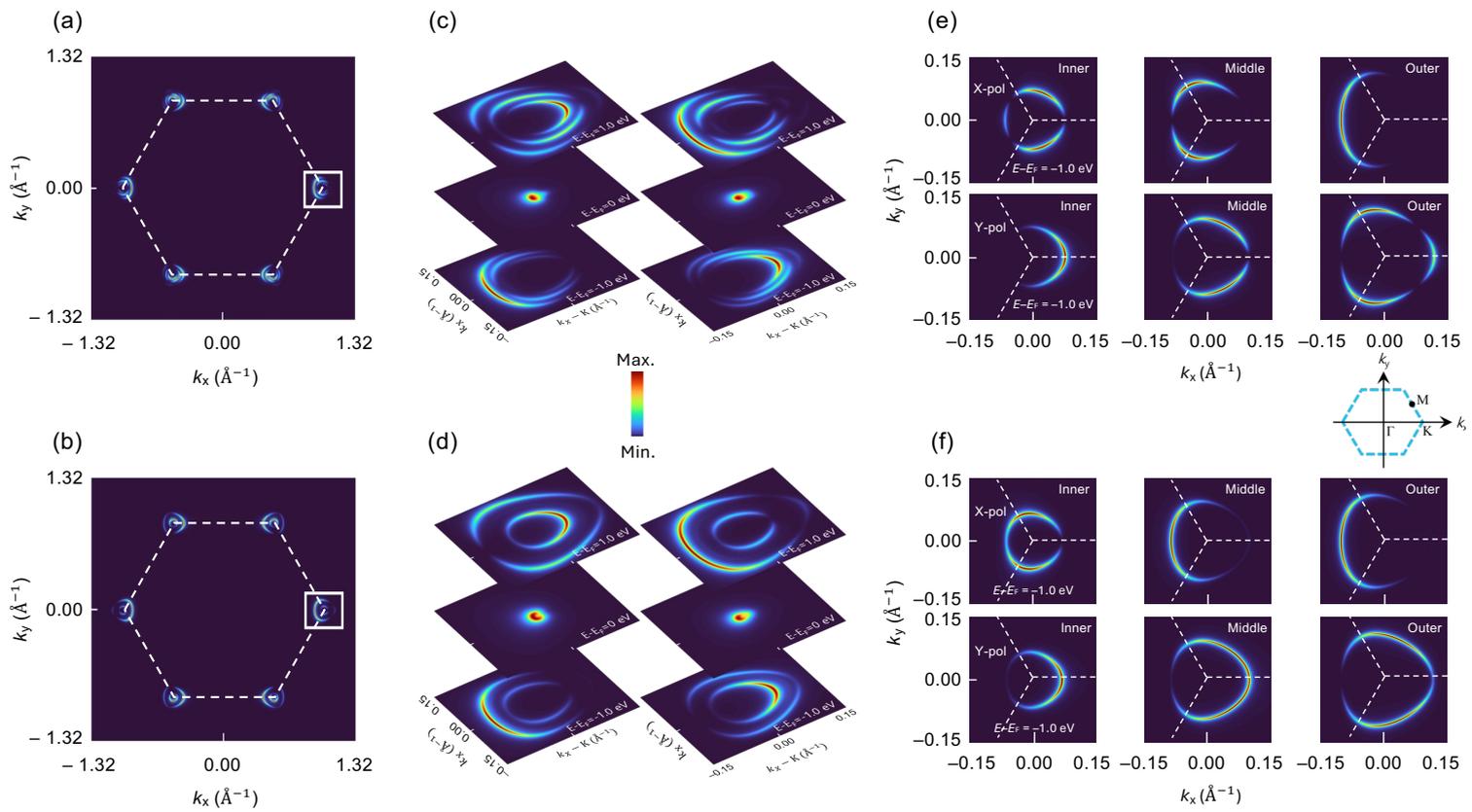

Fig. 2

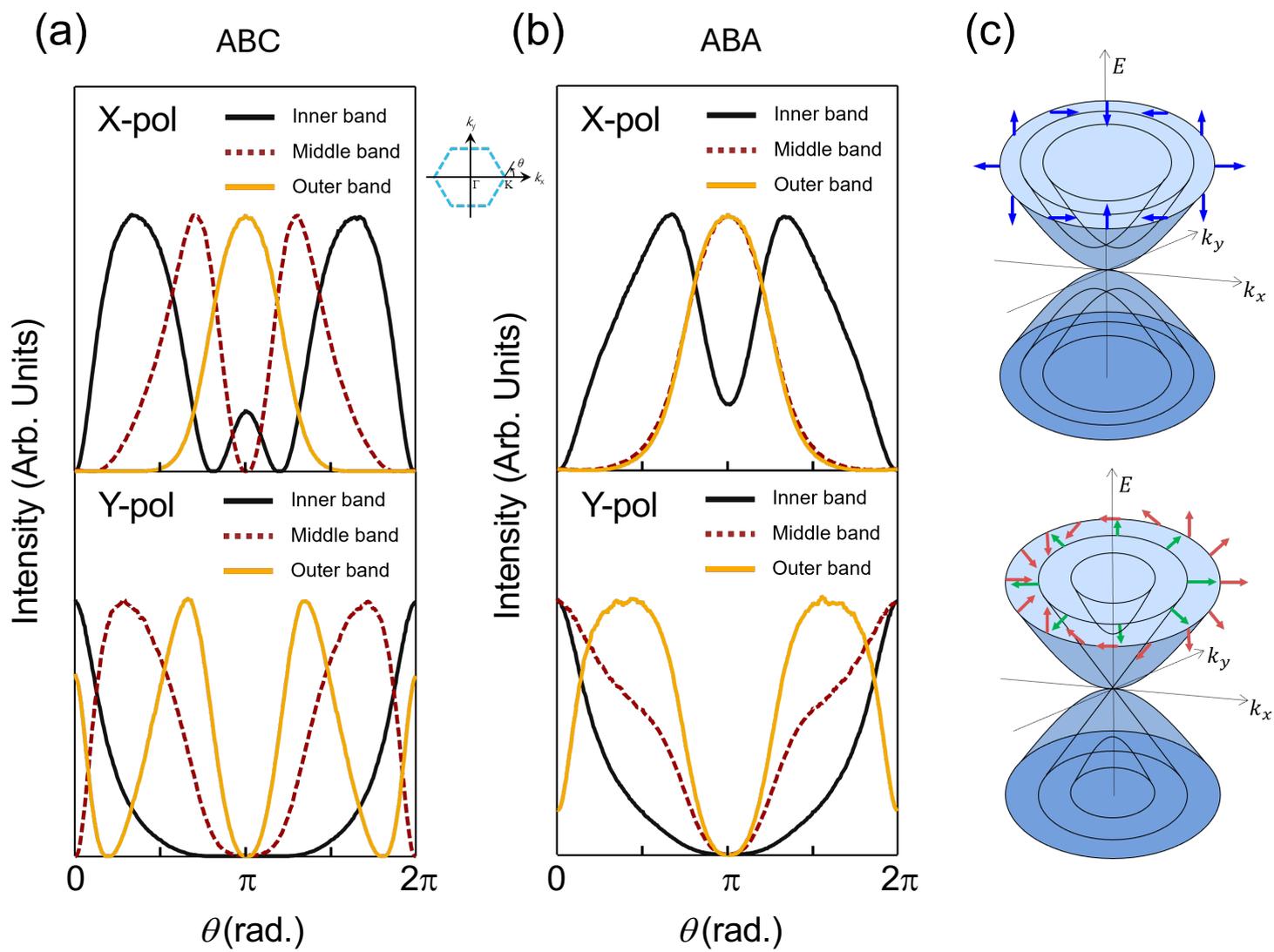

Fig. 3

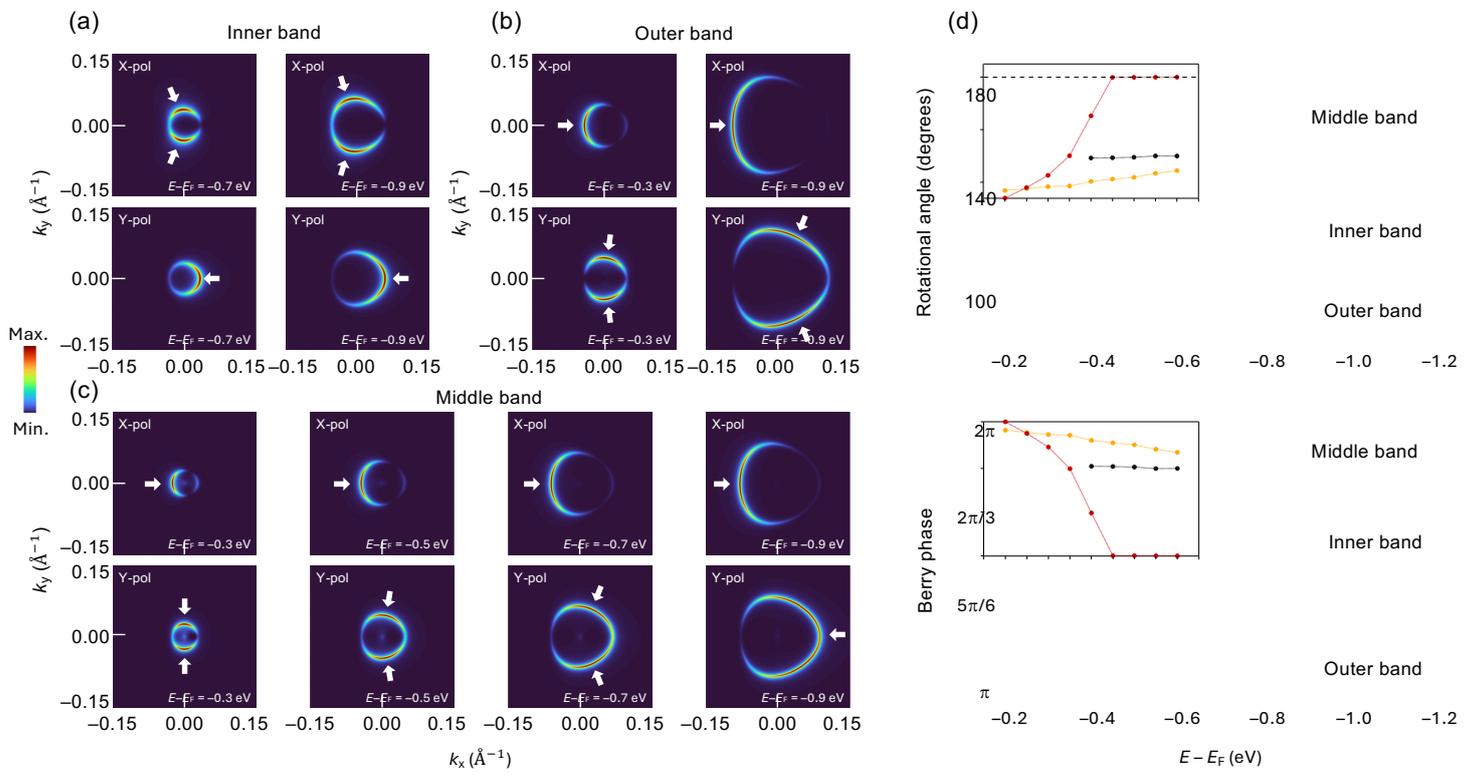

Fig. 4